\documentclass[showpacs,preprintnumbers,amsmath,amssymb]{revtex4}
\usepackage{graphicx}
\usepackage{xcolor}
\begin{document}

\title{Proper relativistic position operators in $1+1$ and $2+1$ dimensions}

\author{Taeseung Choi}

\address{Institute of General Education, Seoul Women's University\\
Seoul 139-774, Korea\\
School of Computational Sciences, Korea Institute for Advanced Study\\
Seoul 130-012, Korea\\
tschoi@swu.ac.kr}





\begin{abstract}

We have revisited the Dirac theory in $1+1$ and $2+1$ dimensions by using the covariant representation of the parity-extended Poincar\'e group in their native dimensions. The parity operator plays a crucial role in deriving wave equations in both theories. We studied two position operators, a canonical one and a covariant one that becomes the particle position operator projected onto the particle subspace.  In $1+1$ dimensions the particle position operator, not the canonical position operator, provides the conserved Lorentz generator. The mass moment defined by the canonical position operator needs an additional unphysical spin-like operator to become the conserved Lorentz generator in $1+1$ dimensions. In $2+1$ dimensions, the sum of the orbital angular momentum given by the canonical position operator and the spin angular momentum becomes a constant of motion. However, orbital and spin angular momentum do not conserve separately. On the other hand the orbital angular momentum given by the particle position operator and its corresponding spin angular momentum become a constant of motion separately.   
  
\keywords{Dirac theory in 1+1 and 2+1 dimensions; ISO(1,2) and ISO(2,1) group representations; Relativistic position operator; Relativistic angular momentum}    
\end{abstract}


\maketitle

\section{Introduction}

The complexity and experimental inaccessibility of the $3+1$ dimensional Dirac theory necessitate the study of the low-dimensional Dirac theory to understand the fundamental nature of relativistic quantum effects. The simulation of a free Dirac fermion has been proposed and implemented using several low-dimensional physical systems \cite{Lurie,Cannata,Lamata,Rusin, Gerritsma,GerritsmaPRL,Salger,Pedernales}. The equivalence of the effective system and the Dirac system is induced by the equality of the Hamiltonian between the two systems. The Hamiltonian alone, however, cannot reveal the subtleties inherent in the theory. Moreover, the Hamiltonian of the low-dimensional Dirac theory has been determined by projection from the physical dimension in general.

One of the fundamental natures of relativistic quantum effect is the Zitterbewegung \cite{Schrodinger}. Experimental results observed in the effective systems have showed Zitterbewegung-like trembling motion \cite{Gerritsma,LeBlanc}, but it is not clear whether these trembling motions are the physical Zitterbewegung or the mere universal interference effects in the two-level system causing self-acceleration \cite{Winkler}. On the contrary, using the quantum field theoretic simulation of the pair-production process in vacuum, Krekora {\it et al.} found that any Zitterbewegung in quantum field theory is prohibited, which means that Zitterbewegung cannot be observed for a real electron \cite{Krekora}. In fact, the existence of the Zitterbewegung for a single Dirac particle was expected depending on the definition of the position operator \cite{Schrodinger,FW,Ourold}. 

That is, the problem of whether the observed trembling motion is the physical Zitterbewegung can be translated into the problem of which position operator is the proper relativistic observable. With this problem in mind, the study of the theory in its native dimension is required. The unitary irreducible representations of the Poincar\'e group in $2+1$ dimensions are given by Binegar \cite{Binegar}. The dynamics determined by the wave equation, however, are represented in the covariant representation, which is non-unitary. The author showed that only unitary representations with half-integer spin for massive particles have relevant massive covariant field theories. They did not discuss the physical operators such as position and spin operators. 

Hence, in this paper, we revisit the $1+1$ and $2+1$ dimensional Dirac theory by directly constructing the covariant representation of the parity-extended Poincar\'e  group in its native dimension. The state vector in the covariant representation is the solution of the Dirac equation that will be shown to be naturally reproduced by using a parity operation on the state vector. The invariance under the parity operation requires the direct sum representation in $1+1$ dimensions similar to the Dirac bispinor in $3+1$ dimensions because there are two inequivalent representations, the left- and the right-handed representations \cite{Ours}. Recently, the change of spin-spin entanglement under Lorentz boosts was studied using the Dirac bispinor structure \cite{Bittencourt}. We have studied two kinds of position operators, one the canonical position operator and the other the covariant position operator in the sense that is defined by the similarity transformation of the canonical position operator with the boost generator, in $1+1$ and $2+1$ dimensions. We shall show that the position operator, which gives the conserved angular momentum in $2+1$ dimensions, is the particle position operator given by the projection of the covariant position operator onto the particle subspace. 
This is similar to $3+1$ dimensional case, where several position operators are discussed through the decomposition of the total angular momentum into orbital angular momentum and spin \cite{Pryce,NW,Fleming64,Choi15}. Hence, the proper spin operator in $3+1$ dimensions would be determined by using the parallel logic from several spin operators \cite{Ours,Bauke,Celeri}. The $1+1$ dimensional case is also interesting because there is no spin in $1+1$ dimensions. In this case the conservation of the Lorentz boost itself will provide the argument for selecting the proper position operator. We will show that the result is the same for $2+1$ dimensional case.       

The paper is organized as follows. In sec. \ref{sec:1DT} we reconstruct the Dirac theory and study two kinds of position operators focusing on the conservation of the Lorentz boost operator in $1+1$ dimensions. In sec. \ref{sec:2DT} we will do similar work in $2+1$ dimensions. In sec. \ref{sec:CON} we conclude with the results.

\section{Proper position operator in the $1+1$ dimensional Dirac theory}
\label{sec:1DT}

\subsection{$1+1$ dimensional Dirac theory}

The symmetry group of the $1+1$ dimensional Dirac theory is the parity-extended inhomogeneous proper orthochronous Lorentz group $ISO(1,1)$. The inhomogeneous Lorentz group $IO(1,1)$ {is defined by the following coordinate transformations $x^{\mu}\rightarrow x^{\prime \mu} $ as
\begin{eqnarray}
(\Lambda,a): x^{\prime \mu}=\Lambda^{\mu}_{\phantom{\mu}\nu} x^\nu +a^\mu
\end{eqnarray}
in the $1+1$ dimensional Minkowski space with the metric tensor $\eta_{\mu\nu}=\mbox{diag}(+,-)$, where $\mu$, $\nu$ $\in \{0,1\}$. $a^\mu$ is an arbitrary spacetime translation vector and the transformations given by $\Lambda^{\mu}_{\phantom{\mu}\nu}$ form a Lorentz group satisfying the conditions
\begin{eqnarray}
\eta_{\mu\nu}\Lambda^{\mu}_{\phantom{\mu}\rho}\Lambda^{\nu}_{\phantom{\mu}\sigma}=\eta_{\rho\sigma}.
\end{eqnarray}
Two successive Poincar\'e transformations on $x^\mu$ give the multiplication law 
\begin{eqnarray}
(\bar{\Lambda},\bar{a})\cdot(\Lambda,a)=(\bar{\Lambda}\Lambda, \bar{\Lambda}a+ \bar{a}),
\end{eqnarray}
which shows that the inhomogeneous Lorentz group is the semi-direct product of the translation and the Lorentz group:
\begin{eqnarray}
IO(1,1)=\mathbb{R}^2 \rtimes O(1,1).
\end{eqnarray}
}

{On physical ground, we are interested in the transformations corresponding to the change of the inertial observers, which is given by the proper orthochronous Lorentz group $ISO(1,1)$ with $\mbox{det}\Lambda=+1$ and $\Lambda^0_{\phantom{0}0} \ge 1$. We call the inhomogeneous proper orthochronous Lorentz group $ISO(1,1)$ $1+1$ dimensional Poincar\'e group.} 
The Lie algebra $iso(1,1)$ of the Poincar\'e group is represented by the translation generators $P^0$ and $P^1$, and the boost generator $J^{01}$. The $iso(1,1)$ algebra is
\begin{eqnarray}
[P^0, P^1]=0, ~~~[ P^\mu, J^{01}]=-i\eta_{\mu 0}P^1 + i \eta^{\mu 1}P^0.
\end{eqnarray} 
The $iso(1,1)$ algebra admits only one second-order Casimir operator 
\begin{eqnarray}
m^2=P^\mu P_\mu,
\end{eqnarray}
which indicates that representations are classified only by the mass $m$. 

There are two inequivalent covariant representations for the Poincar\'e group, whose base states are constructed by acting on the standard momentum state $|m^2;k^\mu\rangle$ {with the Lorentz boost $U(\Lambda)$, i.e.,  
\begin{eqnarray}
\label{eq:VR}
U(\Lambda)|m^2;k^\mu\rangle=e^{i\omega_{01}J^{01}}|m^2;k^\mu\rangle=e^{\pm \xi/2}|m^2;p^\mu\rangle,
\end{eqnarray} 
where $k^\mu=(m,0)$, $p^\mu=\Lambda^\mu_{\phantom{\nu}\nu}k^\nu=(p^0, p^1)$, similar to the $3+1$ dimensional case \cite{Weinberg}. There is no one-dimensional matrix that maps $e^{\xi/2}$ into $e^{-\xi/2}$ by similarity transformation. Here, $|m^2;p^\mu\rangle$ is a general momentum eigenstate and $\omega^{01}=\xi$ is the rapidity of the boost transformation, which is defined by $\tanh^{-1}[|p^1|/(p^0+m)]$. Eq. (\ref{eq:VR}) implies the following conditions
\begin{eqnarray}
\label{eq:LGMM}
J^{01}= -i p^1\partial_{p^0}-i p^0\partial_{p^1}\pm \frac{i}{2}, 
\end{eqnarray}
where $\partial_{p^0}=\partial/(\partial p^0)$ and $\partial_{p^1}=\partial/(\partial p^1)$. }

{$e^{\pm \xi/2}$ are two inequivalent one-dimensional representations of a non-compact $1+1$ dimensional proper orthochronous Lorentz group $SO(1,1)$ and the direction of the boost would be reversed under space inversion;} hence these two representations are transformed to each other under parity (space inversion) as
\begin{eqnarray}
\label{eq:UP11}
\mbox{under parity : } e^{\xi/2} \longleftrightarrow e^{-\xi/2}. 
\end{eqnarray} 
Two inequivalent representations corresponding to $e^{ \xi/2}$ and $e^{- \xi/2}$ are called left- and right-handed representations, respectively, following the $3+1$ dimensional case \cite{Ours}. 

{To construct a representation theory involving parity operation properly, the corresponding representation space requires both the left-handed and the right-handed representation of the Poincar\'e group.} Therefore, the irreducible representation of the $1+1$ dimensional Poincar\'e group extended by parity is obtained by the direct sum representation, whose natural choice of base state is 
\begin{eqnarray}
\label{eq:BST}
\psi_c(p)|m^2;p^\mu\rangle \equiv \left( \begin{array}{c} e^{\xi/2}\\e^{-\xi/2}\end{array}\right) |m^2;p^\mu\rangle.
\end{eqnarray}
The $\psi_c(p)$ is the state in the finite-dimensional representation of the parity-extended $1+1$ dimensional Poincar\'e group depending on the momentum $p^\mu$; hence Eq. (\ref{eq:BST}) is not the tensor product of the two states $\psi(p)$ and $|m^2;p^\mu\rangle$. We will consider only $\psi_c(p)$ because we will work on the momentum representation {with specific momentum $p^\mu$. The state $\psi_c(p)$ is called the chiral representation following $3+1$ dimensional case \cite{Ours}.} 

In this representation, the boost transformation is represented by $\exp{(\sigma_3 \xi/2)}$. Then the state $\psi_c(p)$ can be considered as the boost-transformed state from the rest state $\psi_c(k)$ as
\begin{eqnarray}
\label{eq:CR11}
\psi_c(p)=\frac{1}{\sqrt{2}} e^{\sigma_3 \xi/2} \left( \begin{array}{c} 1 \\ 1\end{array}\right),
\end{eqnarray} 
where   
\begin{eqnarray}
\sigma_1= \left( \begin{array}{cc} 0 & 1 \\ 1 & 0 \end{array}\right), ~~~\sigma_2= \left( \begin{array}{cc} 0 & -i \\ i & 0 \end{array}\right), ~~~\sigma_3=\left( \begin{array}{cc} 1 & 0 \\ 0 & -1 \end{array}\right)
\end{eqnarray}
are three Pauli matrices. 
The state $\psi_c(k)=(1,1)^T/\sqrt{2}$ {with $\xi=0$ in Eq. (\ref{eq:CR11})}, where superscript $T$ denotes transpose, is the rest state because $\xi=0$ in the rest frame. 

The parity operator $\mathcal{P}$ transforms the state $\psi_c(p)$ to 
\begin{eqnarray}
\mathcal{P}\psi_c(p)=\frac{1}{\sqrt{2}}\left( \begin{array}{c} e^{-\xi/2}\\e^{\xi/2}\end{array}\right) 
\end{eqnarray}
from Eq. (\ref{eq:UP11}). 
Then the state $\psi_c(p)$ is normalized by the following Lorentz invariant scalar product as 
\begin{eqnarray}
\label{eq:LISP}
\psi^T_c(p)\mathcal{P}\psi_c(p)=1. 
\end{eqnarray}
The dimension of the irreducible representation of the $1+1$ dimensional Poincar\'e group extended by parity is two; hence another base state orthonormal to the $\psi_c(p)$ under the Lorentz invariant scalar product is needed, which is 
 \begin{eqnarray}
\psi^A_c(p) = \frac{1}{\sqrt{2}}\left( \begin{array}{c} e^{\xi/2}\\-e^{-\xi/2}\end{array}\right),
\end{eqnarray}
and $\psi^T_c(p)\mathcal{P}\psi^A_c(p)=0$, where the meaning of the superscript $A$ will be clear in the following. 

The parity operations on the two base states $\psi_c(p)$ and $\psi^A_c(p)$ are represented by $\sigma_1$ and $-\sigma_1$, respectively. The parity transformed states $\sigma_1\psi_c(p)$ and $-\sigma_1\psi^A_c(p)$ can also be obtained by the action of the standard Lorentz boost $e^{-\xi\sigma_3}$ on each state $\psi_c(p)$ and $\psi^A_c(p)$, respectively. This fact gives the dynamical equations
\begin{eqnarray}
\label{eq:CVDE}
(p^0 \sigma_1 -p^1 \sigma_1\sigma_3-m)\psi_c(p)&=&0, \\ \nonumber
(p^0 \sigma_1 -p^1 \sigma_1\sigma_3+m)\psi^A_c(p)&=&0,
\end{eqnarray}
which correspond to the covariant Dirac equations for a particle and an antiparticle in $1+1$ dimensions similar to the $3+1$ dimensional case \cite{Ours}. Then the superscript $A$ clearly implies an antiparticle. These two covariant Dirac equations can be rewritten as the usual form
\begin{eqnarray}
(p^\mu\gamma_\mu-m)\psi_c(p)=0 \mbox{ and } (p^\mu\gamma_\mu+m)\psi^A_c(p)=0,
\end{eqnarray}
by defining the gamma matrices in the chiral representation 
\begin{eqnarray}
\gamma^0=\sigma_1 \mbox{ and } \gamma^1=-i\sigma_2,
\end{eqnarray}
which satisfy the Clifford algebra 
\begin{eqnarray}
\{\gamma^\mu,\gamma^\nu\}=\eta^{\mu\nu},
\end{eqnarray}
where $\{,\}$ is the anti-commutator. Therefore, $\psi_c(p)$ and $\psi^A_c(p)$ correspond to a particle and an antiparticle state, respectively.  


\subsection{Position operators in $1+1$ dimensions}
\label{sec:PO1}

In this subsection, we will obtain a position operator, which gives the conserved mass moment $J^{01}$ {that should be a conserved quantity, as a consequence of Noether's theorem under the $1+1$ dimensional Poincar\'e symmetry \cite{Noether, Ours}. The translation generators $P^\mu$ become the momenta $p^\mu$.}

For this purpose, we will use the Hamiltonian for a particle in the usual standard representation \cite{Gerritsma}, which is switched from the chiral representation in Eq. (\ref{eq:CVDE}) by the transformation matrix
\begin{eqnarray}
S=\frac{1}{\sqrt{2}}\left( \begin{array}{cc} 1 & 1 \\ 1 & -1 \end{array} \right).
\end{eqnarray} 
The transformation matrix $S$ transforms $\gamma^\mu$ to
\begin{eqnarray}
\gamma^0=\sigma_3, ~~~\gamma^1=i\sigma_2.
\end{eqnarray}
Here we will focus on a particle case because the reasoning for an antiparticle is parallel to the particle case, and straightforward. 
Then in the standard representation, the Hamiltonian of a particle becomes 
\begin{eqnarray}
H= \sigma_1 p^1 + m \sigma_3.
\end{eqnarray}
Then the particle state $\psi(p)$ in the standard representation, which is the eigenstate of the particle Hamiltonian $H$, is obtained by the boost transformation as
\begin{eqnarray}
\label{eq:STHE}
\psi(p)=e^{\sigma_1\xi/2} \left( \begin{array}{c} 1\\  0 \end{array}\right)=e^{\sigma_1\xi/2}\frac{1+\sigma_3}{2} \left( \begin{array}{c} 1\\  0 \end{array}\right).
\end{eqnarray}

We will consider two kinds of position operators: one is the usual canonical position operator $x^0=-i\partial/(\partial p^0)\equiv -i \partial_{p^0}$ and $x^1=i\partial/(\partial p^1)\equiv i\partial_{p^1}$, and the other is the covariant position operator defined by 
\begin{eqnarray}
X^1_N \equiv e^{\sigma_1 \xi/2} x^1 e^{-\sigma_1 \xi/2} \mbox{ and } X^0_N=x^0
\end{eqnarray}
following the $3+1$ dimensional case in Ref. \cite{Ours,Gursey}. Note that both position operators satisfy the canonical commutation relation as
{\begin{eqnarray}
[x^1, p^1]=[X^1_N,p^1]=i.
\end{eqnarray} }

For the canonical position operator, the canonical mass moment $J^{01}_D$ defined as
\begin{eqnarray}
\label{eq:DMM}
J^{01}_D=x^0 p^1 -x^1 p^0,
\end{eqnarray}
is not conserved, because 
{the time derivative of $J^{01}_D$ becomes 
\begin{eqnarray}
\label{eq:MMDP}
\frac{d J^{01}_D}{dt}= -i [J^{01}_D, H]+\frac{\partial J^{01}_D}{\partial t} = -p^0 \sigma_1 +p^1
\end{eqnarray}
in the Heisenberg representation, and also $(dJ^{01}_D/dt) \psi(p)$ is definitely not zero.} Here we use natural units $c=\hbar=1$. Hence the mass moment $J^{01}_D$ corresponding to the canonical position operator is not conserved.

Next let us consider the covariant position operator $X^1_N$, which transforms to the particle position operator $X^1$ as 
\begin{eqnarray}
X^1_N \psi(p) =  e^{\sigma_1 \sigma_3 \xi/2} x^1 e^{-\sigma_1 \sigma_3 \xi/2} \psi(p) \equiv X^1\psi(p),
\end{eqnarray}
when it acts on the particle state $\psi(p)$, using Eq. (\ref{eq:STHE}) and 
\begin{eqnarray}
\exp{(\sigma_1\xi/2)}\frac{1+\sigma_3}{2}= \exp{(\sigma_1\sigma_3 \xi/2)}\frac{1+\sigma_3}{2}.
\end{eqnarray} 
The matrix $e^{-\sigma_1\sigma_3 \xi/2}$ corresponds to the Foldy-Wouthuysen (FW) transformation matrix in $1+1$ dimensions \cite{FW}. Under the similarity transformation by this FW matrix the Hamiltonian transforms to
\begin{eqnarray}
H'=e^{-\sigma_1\sigma_3 \xi/2}H e^{\sigma_1\sigma_3 \xi/2}=E\sigma^3,
\end{eqnarray}
 The particle position operator $X^1$ is explicitly calculated as
\begin{eqnarray}
X^1= x^1 -\frac{m}{2 E^2}\sigma_2.
\end{eqnarray}


The mass moment $J^{01}_P$ corresponding to the particle position operator $X^\mu$ is defined as 
\begin{eqnarray}
J^{01}_P=X^0 p^1 -X^1 p^0.
\end{eqnarray}
This $J^{01}_P$ gives a conserved Noether charge unlike the $J^{01}_D$ in Eq. (\ref{eq:DMM}) because it satisfies
\begin{eqnarray}
\frac{d J^{01}_P}{dt}= -i [J^{01}_P, H]+\frac{\partial J^{01}_P}{\partial t}= - p^1 \frac{p^0 H}{E^2} +p^1 
\end{eqnarray} 
such that
\begin{eqnarray}
\frac{d J^{01}_P}{dt}\psi(p)=0.
\end{eqnarray}
The above considerations imply that the generator of the Lorentz boost $J^{01}$ of the $1+1$ dimensional Poincar\'e group should be given by $J^{01}_P$ corresponding to the particle position operator, not $J^{01}_D$ corresponding to the canonical position operator. 

For completeness, we add a comment on the conservation of the Lorentz boost using the canonical position operator. 
The $J^{01}_D$ alone is not constant of motion as was shown. The spin-like term can be defined by the gamma matrices in the standard representation
as
\begin{eqnarray}
S^{01}_D=\frac{i}{4}[\gamma^0, \gamma^1]=\frac{i}{2}\sigma^1,
\end{eqnarray} 
which has the property
\begin{eqnarray}
[S^{01}_D, H]= i p^0 \sigma^1 -i p^1.
\end{eqnarray}
Hence the sum of $J^{01}_D$ and $S^{01}_D$ gives
\begin{eqnarray}
\frac{d (J^{01}_D + S^{01}_D)}{dt}=-i[ J^{01}_D + S^{01}_D, H]+ \frac{\partial}{\partial t} J^{01}_D=0
\end{eqnarray}
This implies that the sum $J^{01}_D + S^{01}_D$, not $J^{01}_D$ itself, is the representation of the Lorentz generator $J^{01}$. The $J^{01}_D + S^{01}_D$ is nothing but
\begin{eqnarray}
X^0_N p^1 - X^1_N p^0
\end{eqnarray}
and {is also the same as the direct sum representation of the Lorentz generators in Eq. (\ref{eq:LGMM}).} 
However, there is no physical operator corresponding to the $S^{01}_D$ in $1+1$ dimensions. Consequently we can conclude that the particle position operator is the proper position operator for $1+1$ dimensional Dirac theory. 

\section{Proper position operator in the $2+1$ dimensional Dirac theory}
\label{sec:2DT}

\subsection{$2+1$ dimensional Dirac theory}
The Poincar\'e algebra $iso(2,1)$ can be obtained by projection from $3+1$ dimensional Poincar\'e algebra \cite{Weinberg}. There are two Casimir operators in the $2+1$ dimensional Poincar\'e group $ISO(2,1)$; hence all unitary irreducible representations of the $2+1$ dimensional Poincar\'e group are given by $(m,s)$, where the real number $s$ labels the irreducible representation of the little group $SO(2)$ in the form of $\exp{(i s\theta)}$ for the rotation angle $\theta$ \cite{Binegar,Bekaert}.  

Here we are interested in the covariant representation for spin $1/2$ because the covariant state for spin $1/2$ satisfies the $2+1$ dimensional Dirac equation \cite{Binegar}. The covariant state for spin $1/2$ can be denoted as 
\begin{eqnarray}
\psi(p)|m^2;p^\mu\rangle
\end{eqnarray}
where $p^\mu=(p^0, p^1, p^2)$ and $\psi(p)$ is the spinor state that will be explicitly determined in the following. 

The faithful irreducible spinor representation of the Lorentz generator for the $2+1$ dimensional Poincar\'e group is given by
\begin{eqnarray}
\label{eq:LG12}
J^{01}&=&\frac{i}{2}\sigma^1,~~~J^{02}=\frac{i}{2}\sigma^2, ~~~J^{12}=\frac{\sigma^3}{2},
\end{eqnarray} 
through the Clifford algebra 
\begin{eqnarray}
\gamma^\mu \gamma^\nu +\gamma^\nu \gamma^\mu=\eta^{\mu\nu}
\end{eqnarray}
with
\begin{eqnarray}
\gamma^0 &=&\sigma^3, ~~~\gamma^1=i\sigma^2, ~~~\gamma^2=-i\sigma^1
\end{eqnarray}
in the standard representation 
similar to the $3+1$ dimensional case \cite{Ours}. Here the metric tensor $\eta^{\mu\nu}=\mbox{diag}(+,-,-)$. 
Then the Lorentz transformation is represented in the spinor space by 
\begin{eqnarray}
\label{eq:LG}
e^{iJ^{\mu\nu}\omega_{\mu\nu}/2}=e^{ -iJ^{0i}\omega^{0i}+i J^{12}\omega^{12}}= e^{\pm \boldsymbol{\sigma}\cdot{\boldsymbol{\xi}}/2+i\sigma^3 \theta/2}
\end{eqnarray}
where $\boldsymbol{\sigma}\cdot{\boldsymbol{\xi}}=\sigma^k \xi^k$, $\omega^{0i}=\xi^i$ is the $i$-component of the rapidity vector $\boldsymbol{\xi}$ with $\sqrt{\boldsymbol{\xi}\cdot\boldsymbol{\xi}}=\tanh^{-1}[\sqrt{{\bf p}\cdot {\bf p}}/(p^0+m)]$, and $\omega^{12}$ is the rotation angle $\theta$. We use the Einstein summation convention. The Latin indices run through $1$ and $2$ and the Greek indices run from $0$ to $2$.

In Eq. (\ref{eq:LG}) the $\pm$ correspond to the left-handed and the right-handed representation of the Lorentz boost. {Note that the left-handed representation is equivalent to the right-handed representation because they are transformed into each other by the similarity transformation with $\sigma^3$. Therefore the spinor representation of the parity-extended $2+1$ dimensional Poincar\'e group given by $\psi^{P/AP}(p)$ is equivalent to the spinor representation of the $2+1$ dimensional Poincar\'e group, which is different from the $1+1$- and $3+1$ dimensional cases.} The two spin eigenstates for $\sigma^3/2$ constitute the base states of the fundamental spin state representation of $SO(2,1)$, which are
\begin{eqnarray}
\psi^P(k)=\left( \begin{array}{c} 1 \\ 0 \end{array}\right),~~ \psi^{AP}(k)=\left( \begin{array}{c} 0 \\ 1 \end{array}\right)
\end{eqnarray} 
with $k^\mu=(m,0,0)$. Then the boosted state $\psi^{P/AP}(p)$ with $p^\mu=(p^0,p^1,p^2)$ becomes
\begin{eqnarray}
\psi^{P/AP}(p) =  e^{\pm \boldsymbol{\sigma}\cdot{\boldsymbol{\xi}}/2}\psi^{P/AP}(k). 
\end{eqnarray}

The parity operator is represented by $\sigma^3$ and $-\sigma^3$ for $\psi^P(p)$ and $\psi^{AP}(p)$, respectively, where the $-$ sign in $-\sigma^3$ is determined based on the fact that the parity operation does not change the state at the rest frame. The parity operations $\sigma^3$ and $-\sigma^3$ on $\psi^P(p)$ and $\psi^{AP}(p)$, respectively, derive the covariant Dirac equations for a particle and antiparticle
\begin{subequations}
\begin{eqnarray}
\label{eq:CDE}
(\gamma^\mu p_\mu - m)\psi^P(p) &=& 0 \\
(\gamma^\mu p_\mu + m)\psi^{AP}(p) &=& 0.
\end{eqnarray}
\end{subequations}
Hence, the superscripts $P$ and $AP$ denote particle and antiparticle, respectively. 

\subsection{Position operator in $2+1$ dimensions}

As in the $1+1$ dimensional case in sec. \ref{sec:PO1}, we will consider the two kinds of position operators: the usual canonical position operators $x^0=i\partial_{p^0}$ and $x^k=-i \partial_{p^k}$, and the covariant position operators defined by
\begin{eqnarray}
X^k_N=e^{ \boldsymbol{\sigma}\cdot{\boldsymbol{\xi}}/2} x^k e^{ -\boldsymbol{\sigma}\cdot{\boldsymbol{\xi}}/2} \mbox{ and } X^0_N=x^0.
\end{eqnarray}
Both kinds of position operators satisfy the canonical relations, i.e., 
\begin{eqnarray}
[x^k,p^l]=[X^k_N, p^l]=i\delta_{kl},
\end{eqnarray} 
where $\delta_{kl}$ is the Kronecker delta.

In the $2+1$ dimensional case, there is a spin that labels the representation of the little group $SO(2)$. The 2-dimensional spinor $\psi(p)$ is the direct sum of the particle spinor $\psi^P(p)$ and the antiparticle spinor $\psi^{AP}(p)$. The spin operator for this 2-dimensional representation is
\begin{eqnarray}
\frac{\sigma^3}{2}
\end{eqnarray}
given by the usual definition of the spin operator as
\begin{eqnarray}
S_D^{12}=\frac{i}{4}[\gamma^1, \gamma^2],
\end{eqnarray}
where $S_D^{12}$ is the spin angular momentum that gives the total angular momentum 
\begin{eqnarray}
\label{eq:TAM}
J^{12}_D=S_D^{12}+L_D^{12}
\end{eqnarray}
by addition of the canonical orbital angular momentum defined by
\begin{eqnarray}
L_D^{12}=x^1 p^2-x^2 p^1.
\end{eqnarray}

The Dirac Hamiltonian for a particle becomes
\begin{eqnarray}
H= \boldsymbol{\sigma}\cdot {\bf p}+m \sigma^3. 
\end{eqnarray}
from Eq. (\ref{eq:CDE}). One can easily check that the total angular momentum 
$J^{12}_D$
is conserved, but the orbital and the spin angular momentum are not conserved separately., i.e.,  
\begin{eqnarray}
[L_D^{12}, H]= i\sigma^1 p^2 - i\sigma^2 p^1 \mbox{ and } [S_D^{12}, H]=-i\sigma^1 p^2 + i\sigma^2 p^1.
\end{eqnarray} 

Next, let us consider the conservation of the angular momentum corresponding to the covariant position operator $X^k_N$. $X^k_N$ becomes the particle position operator $X^k$ when it acts on the particle spinor $\psi^P(p)$ as follows
\begin{eqnarray}
X^k_N \psi^P(p)&=& X^k \psi^P(p)= e^{-\sigma^3 \boldsymbol{\sigma}\cdot{\boldsymbol{\xi}}/2}x^k e^{ \sigma^3 \boldsymbol{\sigma}\cdot{\boldsymbol{\xi}}/2}\psi^P(p) \\ \nonumber
&=& \left[ x^k -ip^k\frac{E+\sigma^3 \boldsymbol{\sigma}\cdot{\bf p}}{2E^2(E+m)} +i \frac{(E+m)\sigma^3 \sigma^k+\boldsymbol{\sigma}\cdot{\bf p}\sigma^k}{2E(E+m)} \right] \psi^P(p),
\end{eqnarray}
where the unitary matrix 
\begin{eqnarray}
e^{\sigma^3 \boldsymbol{\sigma}\cdot{\boldsymbol{\xi}}/2}=\frac{E+m +\sigma^3 \boldsymbol{\sigma}\cdot{\bf p}}{\sqrt{2E(E+m)}}
\end{eqnarray}
 corresponds to the unitary FW transformation matrix. 
And the corresponding particle spin is defined by
\begin{eqnarray}
S^{12}= e^{-\sigma^3 \boldsymbol{\sigma}\cdot{\boldsymbol{\xi}}/2} \frac{\sigma^3}{2} e^{\sigma^3 \boldsymbol{\sigma}\cdot{\boldsymbol{\xi}}/2}=\frac{I}{2},
\end{eqnarray}
where $I$ is the two-dimensional identity matrix. One can easily check that the particle orbital angular momentum operator, 
\begin{eqnarray}
L^{12}=X^1p^2-X^2p^1,
\end{eqnarray}
and the particle spin operator commute with the Dirac Hamiltonian $H$ separately. 
The particle total angular momentum defined as
\begin{eqnarray}
J^{12}=L^{12}+S^{12}=X^1p^2-X^2p^1 + \frac{I}{2}
\end{eqnarray}
is {equal to the total angular momentum $J^{12}_D$ in Eq. (\ref{eq:TAM}), which is the generator of the spatial rotation defined by $ip^2\partial_{p^1}-ip^1\partial_{p^2}+i\sigma^1/2$. That is, both total angular momentum operators corresponding to the canonical and the covariant position operator are the same as the rotation generator of the $2+1$ dimensional Poincar\'e group.} The little group symmetry requires that the spin should be a conserved quantity as a consequence of Noether's theorem \cite{Noether, Ours}. This fact implies that the position operator corresponding to the spin operator, which transforms the spinor under the little group symmetry, is the covariant and particle position operators. 

The mass moment operator $J^{0k}$ in $iso(2,1)$ algebra has no spin part; however, to obtain the conserved mass moment the following spin-like term 
\begin{eqnarray}
S^{0k}_c=\frac{i}{4}[\gamma^0,\gamma^k]=\frac{i}{2}\sigma^k
\end{eqnarray}   
should be added to the canonical mass moment of 
\begin{eqnarray}
J^{0k}_c=x^0 p^k -x^k p^0
\end{eqnarray}
similar to the mass moment in $iso(1,1)$ algebra. However, the mass moment $J^{0k}$ does not play any crucial role to determine a proper position operator, unlike the total angular momentum $J^{12}$, which has the actual spin operator term.

\section{Conclusions}
\label{sec:CON}

We studied the Dirac theory and two kinds of position operators, one the usual canonical position operator and the other the covariant position operator, in $1+1$ and $2+1$ dimensions. In $1+1$ dimensions, there exist two inequivalent representations, the left-handed and the right-handed representations, which are transformed into each other under the parity operation. Using the direct-sum representation for the parity-extended $ISO(1,1)$ group and the parity operation, we derived the dynamical equations for a particle and an antiparticle in $1+1$ dimensional Dirac theory. To represent the conserved Lorentz generator $J^{01}$ of $ISO(1,1)$ group, the mass moment operator given by the canonical position operator needs an additional spin-like operator. However, the covariant position operator, defined by the similarity transformation of the canonical position operator with the boost generator, provides the conserved mass moment operator through the particle position operator without requiring the artificial spin-like term. This suggests that the proper position operator is the covariant position operator because there is no physical spin-like operator to transform the internal space in $1+1$ dimensions. 

In $2+1$ dimensions, the faithful two-dimensional spinor representation provides equivalent representation between the left-handed and the right-handed representations. Hence, the two-dimensional spinor representation includes both the left-handed and the right-handed spinors. We also derived the covariant dynamical equations for a particle and an antiparticle using a parity operation in $2+1$ dimensional Dirac theory. We have shown that the covariant position operator equally defined as in $1+1$ dimensions becomes the particle position operator acting on the particle spinor and its corresponding orbital and spin angular momentum are conserved separately. As a result, we conclude that the covariant position operator is also the proper position operator in $2+1$ dimensional Dirac theory, because the spin operator should be a constant of motion reflecting the fact that the spin operator is the generator of the little group symmetry.

\section*{Acknowledgements}

 This work was supported by the Basic Science Research Program through the National Research Foundation of
Korea (NRF) funded by the Ministry of Education (2018-0239) and by a research grant from Seoul Women’s University(2020-0278). T. Choi thanks Y.D. Han for helpful discussions.



\section*{References}

\begin{thebibliography}{0}

\bibitem{Lurie} D. Lurié and S. Cremer, \textit{Physica}  {\bf 50} 224 (1970). 
\bibitem{Cannata} F. Cannata, L. Ferrari and G. Russo, \textit{Solid State Commun.} {\bf 74} 309 1990. 
\bibitem{Lamata} L. Lamata, J. León, T. Schätz and E. Solano, \textit{Phys. Rev. Lett.} {\bf 98}, 253005 (2007). 
\bibitem{Rusin} T. M. Rusin and W. Zawadzki, \textit{Phys. Rev. B} {\bf 78}, 125419 (2008). 
\bibitem{Gerritsma} R. Gerritsma, G. Kirchmair, F. Zahringer, E. Solano, R. Blatt and C. Roos, \textit{Nature} {\bf 463} 68 (2010).
\bibitem{GerritsmaPRL} R. Gerritsma \textit{et al.}, \textit{Phys. Rev. Lett.} {\bf 106}, 060503 (2011).
\bibitem{Salger}  T. Salger, C. Grossert, S. Kling and M. Weitz, \textit{Phys. Rev. Lett.} {\bf 107}, 240401 (2011). 
\bibitem{Pedernales}  J. S. Pedernales \textit{et al.}, \textit{Sci. Rep.} {\bf 5}, 15472 (2015). 


\bibitem{Schrodinger}
 Schr\"odinger E.,
 \textit{Sitz. Preuss. Akad. Wiss. Phys.-Math. Kl.}
 \textbf{24}, 418-428 (1930).

\bibitem{LeBlanc} L. J. LeBlanc \textit{et al.}, \textit{New J. Phys.} {\bf 15}, 073011 (2013).
\bibitem{Winkler} R. Winkler, U. Z\"ulicke and J. Bolte, \textit{Phys. Rev. B} {\bf 75}, 205314 (2007).
\bibitem{Krekora} P. Krekora, Q. Su, and R. Grobe, \textit{Phys. Rev. Lett.} {\bf 93}, 043004 (2004). 
\bibitem{FW} L. L. Foldy and S. A. Wouthuysen, \textit{Phys. Rev.} {\bf 78}, 29 (1950).
\bibitem{Ourold}  T. Choi and S. Y. Cho, \textit{Spin Operators for Massive Particles}, arXiv:1410.0468[quant-ph].
\bibitem{Binegar} B. Binegar, \textit{J. Math. Phys.} {\bf 23}, 1511 (1981).
\bibitem{Ours} T. Choi and S. Y. Cho, \textit{Spin operators and representations of the Poincaré group}, arXiv:1807.06425[physics.gen-ph].
\bibitem{Bittencourt} V. A. S, V. Bittencourt, A. E. Bernardini, and M. Blasone, \textit{Phys. Rev. A} {\bf 97}, 032106 (2018). 
\bibitem{Pryce} M. H. L. Pryce, \textit{Proc. R. Soc. Lond. A} {\bf 195}, 62 (1948). 
 %
 \bibitem{NW} T. D. Newton and E. P. Wigner, \textit{Rev. Mod. Phys.} {\bf 21}, 400 (1949).
\bibitem{Fleming64} 
G. N. Fleming, 
\textit{Phys. Rev.} \textbf{137}, (1964) B188. 
\bibitem{Choi15} T. Choi, \textit{J. Korean Phys. Soc.} {\bf 66}, 877 (2015).
\bibitem{Bauke} H. Bauke, S. Ahrens, C. H. Keitel, and R. Grobe, \textit{New J. Phys.} {\bf 16}, 043012 (2014).
\bibitem{Celeri} L. C. C\'eleri, V. Kiosses, and D. R. Terno, \textit{Phys. Rev. A} {\bf 94}, 062115 (2016).
\bibitem{Weinberg} 
S. Weinberg, \textit{The Quantum theory of fields}, Cambdrdge University Press, New York, U.S.A. (2005).
\bibitem{Noether} 
E. Noether, \textit{Nachr. Ges. Wiss. G¨ott., Math.
Phys. Kl.} \textbf{II} (1918) 235; English translation by M. A. Travel, \textit{Transport Theory and Statistical Physics} \textbf{1} (1971) 183. 
\bibitem{Gursey}
  F. G\"ursey,
 \textit{Phys. Lett.} \textbf{14} (1965) 330.

\bibitem{Bekaert} X. Bekaert and N. Boulanger, \textit{The unitary representations of the Poincare group in any spacetime dimension}, arXiv:hep-th/0611263.
 






\end{thebibliography}

\end{document}